\documentclass[twocolumn,preprintnumbers,nofootinbib]{revtex4-1}
\usepackage[english]{babel}
\usepackage[utf8x]{inputenc}
\usepackage{amssymb}
\usepackage{amsmath}
\usepackage{graphicx}

\usepackage{xcolor}
\definecolor{lcolor}{rgb}{0.5,0,0}
\definecolor{citcolor}{rgb}{0,0.3,0.0}

\usepackage[breaklinks,colorlinks,urlcolor=blue,citecolor=citcolor,linkcolor=lcolor]{hyperref}

\newcommand{\ud}{\, \mathrm{d}}
\newcommand{\xt}{{{\boldsymbol x}_\perp}}
\newcommand{\yt}{{{\boldsymbol y}_\perp}}
\newcommand{\bt}{{{\boldsymbol b}_\perp}}
\newcommand{\rt}{{{\boldsymbol r}_\perp}}
\newcommand{\tr}{\, \mathrm{Tr} \, }
\def\P{{\boldsymbol P}}
\newcommand{\Pt}{{\P_\perp}}
\newcommand{\qs}{Q_\mathrm{s}}
\newcommand{\qso}{Q_\mathrm{s0}}
\newcommand{\lqcd}{\Lambda_{\mathrm{QCD}}}
\newcommand{\as}{\alpha_{\mathrm{s}}}

\begin{document}

\author{B. Ducloué}
\affiliation{University	of Jyväskylä, Department of Physics, P.O. Box 35, FI-40014 University of Jyväskylä, Finland}
\affiliation{Helsinki Institute of Physics, P.O. Box 64, FI-00014 University of Helsinki, Finland}

\title{Nuclear modification of forward Drell-Yan production at the LHC}

\begin{abstract}
Forward Drell-Yan production at high energy can provide important constraints on gluon densities at small $x$, in the saturation regime. In this work we focus on the nuclear modification of this process, which could be measured at the LHC in the near future. For this we employ the color dipole approach, using the optical Glauber model to relate the dipole cross section of a nucleus to the one of a proton. Combining these results with our earlier results for forward $J/\psi$ production, we compute the ratio of the nuclear modification factors of these two processes. This observable was recently suggested as a way to distinguish between initial and final state effects in forward particle production.

\end{abstract}

\maketitle

\section{Introduction}

Particle production at forward rapidity in high energy proton-proton and proton-nucleus collisions has been the subject of numerous studies aiming at improving our understanding of saturation dynamics. Indeed, these processes probe the target proton or nucleus at very small $x$ which is where saturation effects should be enhanced. Two important examples of such processes are light hadron and quarkonium production, for which it was shown that saturation could provide an explanation for the nuclear suppression observed at the LHC~\cite{Lappi:2013zma,Ducloue:2015gfa,Ma:2015sia,Ducloue:2016pqr}. However, these processes are also sensitive to fragmentation and final state effects. In this respect, Drell-Yan production appears as a much cleaner probe of initial state effects in hadronic collisions. In particular, it was recently suggested that the ratio of the nuclear modification factors of $J/\psi$ and Drell-Yan production could be used as a way to distinguish between various approaches, based on either initial or final state effects, that can describe the rapidity dependence of the nuclear modification of forward $J/\psi$ production at the LHC~\cite{Arleo:2015qiv}. Therefore, one of the main motivations of the present work is to make predictions for this observable in the saturation approach. For this, we will first study the nuclear modification of forward Drell-Yan production at the LHC using the dipole correlators introduced in Ref.~\cite{Lappi:2013zma}. The comparison of these results with future measurements of this observable would provide an additional test for these correlators which have been shown to lead to a rather good agreement with experimental data on the nuclear modification of single inclusive forward hadron~\cite{Lappi:2013zma} and $J/\psi$~\cite{Ducloue:2015gfa,Ducloue:2016pqr} production. Such a measurement could be performed at the ALICE or LHCb experiments at the LHC in the near future.

\section{Formalism}

\begin{figure}
	\centering
	\includegraphics[scale=0.6]{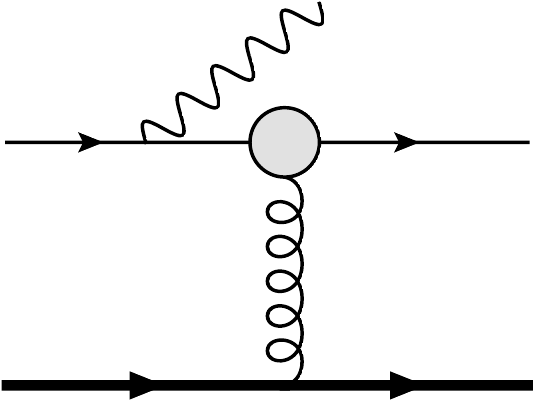}
	\hspace{1cm}
	\includegraphics[scale=0.6]{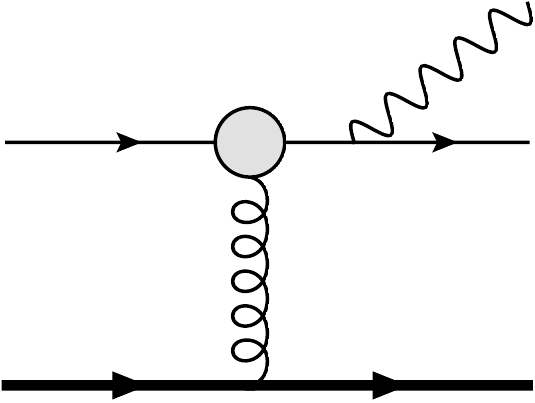}
	\caption{Diagrams contributing to Drell-Yan production in the color dipole approach.}
	\label{fig:diags_sat}
\end{figure}

\begin{figure}
	\centering
	\includegraphics[scale=0.55]{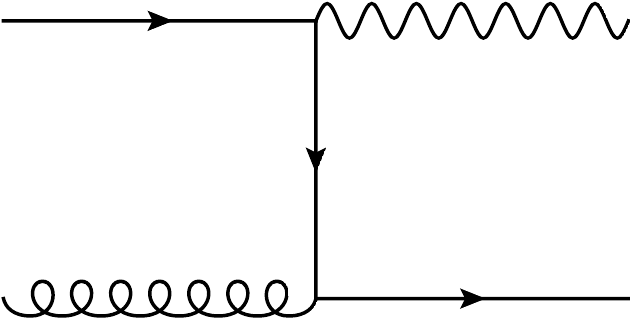}
	\hspace{1cm}
	\includegraphics[scale=0.55]{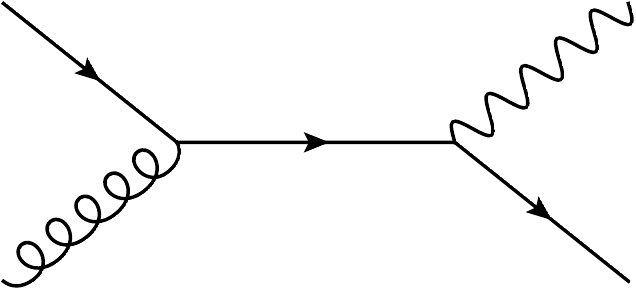}
	\caption{Lowest order contributions probing the gluon density of the target in collinear factorization.}
	\label{fig:diags_coll}
\end{figure}

The study of the Drell-Yan process in the color dipole approach has been the subject of many theoretical and phenomenological works, see for example Refs.~\cite{Brodsky:1996nj,Kopeliovich:2000fb,Kopeliovich:2001hf,Gelis:2002fw,Gelis:2006hy,GolecBiernat:2010de,Stasto:2012ru,Ducati:2013cga,Motyka:2014lya,Basso:2015pba,Schafer:2016qmk,Basso:2016ulb,Motyka:2016lta,Brzeminski:2016lwh}. In this formalism, the physical picture is the following: a collinear quark emitted by the projectile proton can radiate a virtual photon either before or after interacting with the dense color field of the target. This virtual photon then decays into a dilepton pair. These two contributions are shown in Fig.~\ref{fig:diags_sat}. In collinear factorization, contributions involving explicitly the target's gluon density start to appear only at next-to-leading order (see Fig.~\ref{fig:diags_coll}, which represents a subset of the contributions included in Fig.~\ref{fig:diags_sat}). In the kinematics considered here these contributions are enhanced by the strong rise of gluon densities at small $x$. Using similar notations as in Ref.~\cite{Ducati:2013cga}, the dilepton pair production cross section can be written, in the limit of massless quarks, as
\begin{widetext}
\begin{align}\label{eq:dsigma}
\frac{\ud\sigma}{\ud Y \! \ud M^2 \! \ud^2 \! \Pt \! \ud^2 \bt}  = 
\frac{\alpha_{em}^2}{3\pi^3M^2}\int_{x_{1}}^{1} \!\! \frac{\ud\alpha}{\alpha}
\sum_f e_f^2 \left[ q_f \left(\frac{x_{1}}{\alpha},Q^2\right) + \bar{q}_f \left(\frac{x_{1}}{\alpha},Q^2\right) \right] \!
\bigg[ & 2M^2(1-\alpha)^2 \left(\frac{{\cal I}_1}{P_\perp^2+\varepsilon^2}-\frac{ {\cal I}_2}{4\varepsilon} \right)  \\ \nonumber
& + [1+(1-\alpha)^2]\left( \frac{\varepsilon \, P_\perp \, {\cal I}_3}{P_\perp^2+\varepsilon^2} -\frac{{\cal I}_1}{2}+\frac{\varepsilon \,{\cal I}_2}{4}\right) \! \bigg] ,
\end{align}
\end{widetext}
where $Y$, $M$ and $\Pt$ are respectively the rapidity, invariant mass and transverse momentum of the dilepton pair, $x_1=\sqrt{P_\perp^2+M^2} \, e^Y/\sqrt{s}$, $\bt$ is the target's impact parameter and $\varepsilon^2 = (1-\alpha)M^2$. $\mathcal{I}_1$, $\mathcal{I}_2$ and $\mathcal{I}_3$  read
\begin{eqnarray}
\mathcal{I}_1 & = & \int_0^{\infty} \!\!\!\! \ud r \, r \, J_0(P_\perp r) K_0(\varepsilon r) \mathcal{N}_{x_2}(\alpha r, \bt) \, , \nonumber \\
\mathcal{I}_2 & = & \int_0^{\infty} \!\!\!\! \ud r \, r^2 \, J_0(P_\perp r) K_1(\varepsilon r) \mathcal{N}_{x_2}(\alpha r, \bt) \, ,\nonumber \\
\mathcal{I}_3 & = & \int_0^{\infty} \!\!\!\! \ud r \, r \, J_1(P_\perp r) K_1(\varepsilon r) \mathcal{N}_{x_2}(\alpha r, \bt) \, .
\label{eq:I123}
\end{eqnarray}
The description of the projectile proton in terms of collinear quark distributions $q_f$ is justified by the fact that the longitudinal momentum fraction $x_1/\alpha$ at which it is probed is not very small at forward rapidity. On the other hand, the target is probed at very small $x_2$ and it can thus be described in terms of classical color fields. The information about its gluon density is contained in the dipole scattering amplitude $\mathcal{N}$, which is related to $S$, the fundamental representation dipole correlator in the color field of the target:
\begin{align}
\mathcal{N}(\rt=\xt-\yt)&=1-S(\xt-\yt) \\ \nonumber
&=1-\frac{1}{N_c}\left< \tr U^\dag(\xt)U(\yt)\right>,
\end{align}
where $U(\xt)$ is a fundamental representation Wilson line in the color field of the target. In Ref.~\cite{Ducati:2013cga}, $x_2$ is taken as 
\begin{equation}
x_2=\frac{\sqrt{P_\perp^2+M^2}}{\sqrt{s}} \, e^{-Y} \equiv x_2^\text{min} \; .
\label{eq:x2_min}
\end{equation}
A more detailed treatment of the kinematics taking into account the unobserved outgoing quark leads to
\begin{equation}
x_2=\frac{\sqrt{P_\perp^2+M^2}}{\sqrt{s}} \, e^{-Y} \left(1 + \frac{\alpha}{1-\alpha} \, \frac{q_\perp^2}{P_\perp^2+M^2} \right) \; ,
\label{eq:x2_exact}
\end{equation}
where $q_\perp$ is the transverse momentum of the outgoing quark. Therefore~(\ref{eq:x2_min}) is strictly speaking the minimal value allowed for $x_2$. On the other hand, it is not possible to use~(\ref{eq:x2_exact}) directly since $q_\perp$ has already been integrated over to arrive at Eq.~(\ref{eq:dsigma}). To estimate the importance of the choice of $x_2$, we will use both~(\ref{eq:x2_min}) and an effective value
\begin{equation}
x_2=\frac{\sqrt{P_\perp^2+M^2}}{\sqrt{s}} \, e^{-Y} \left(1 + \frac{\alpha}{1-\alpha} \, \frac{P_\perp^2+\qs^2}{P_\perp^2+M^2} \right) \equiv x_2^\text{eff} \; ,
\label{eq:x2_eff}
\end{equation}
where $\qs$ is the saturation scale of the target (we use the same definition of the saturation scale as in Ref.~\cite{Lappi:2013zma}, i.e. $\qs$ is defined as the solution of $\mathcal{N}(r_\perp^2=2/\qs^2)=1-e^{-1/2}$). The expression in Eq.~(\ref{eq:x2_eff}) is motivated by the fact that on average the total transverse momentum provided by the target should be of the order of its saturation scale.

In this work we use the dipole correlators introduced in Ref.~\cite{Lappi:2013zma}. The rapidity (or $x$) evolution of $S$ is obtained by solving numerically the Balitsky-Kovchegov equation with running coupling corrections~\cite{Balitsky:1995ub,Kovchegov:1999ua,Balitsky:2006wa}. In the case of a proton target, the initial condition at $x_0=0.01$ is parametrized as
\begin{equation}\label{eq:icp}
S^\text{p}_{x_0}(\rt) = \exp \left[ -\frac{\rt^2 \qso^2}{4} \ln \left(\frac{1}{|\rt| \lqcd}\!+\!e_c \cdot e\right)\right] ,
\end{equation}
and it is assumed that there is no impact parameter dependence in $S$, therefore when computing proton-proton cross sections we make the replacement \begin{equation}\label{eq:defsigma0}
\int \ud^2 \bt \to \frac{\sigma_0}{2} \; ,
\end{equation}
where $\sigma_0/2$ is the effective proton transverse area. The running coupling in coordinate space is taken as
\begin{equation}
\as(r) = \frac{12\pi}{(33 - 2N_f) \log \left(\frac{4C^2}{r^2\lqcd^2} \right)} \, .
\end{equation}
A fit of the free parameters in these expressions to HERA DIS data~\cite{Aaron:2009aa} leads to $\qso^2= 0.060$ GeV$^2$, $C^2= 7.2$, $e_c=18.9$ and $\sigma_0/2 = 16.36$ mb~\cite{Lappi:2013zma}. Because of the lack of accurate nuclear DIS data at small $x$ a similar fit cannot be performed for a nuclear target. To extrapolate the proton dipole correlator to a nucleus, we use, as in Ref.~\cite{Lappi:2013zma}, the optical Glauber model. In this model, the probe coming from the projectile proton is supposed to scatter independently off the target nucleons at the initial rapidity and, after averaging over the fluctuating positions of the nucleons in the nucleus, we get
\begin{multline}\label{eq:ica}
S^\text{A}_{x_0}(\rt,\bt) = \exp\bigg[ -A T_A(\bt) 
\frac{\sigma_0}{2} \frac{\rt^2 \qso^2}{4} 
\\ \times
\ln \left(\frac{1}{|\rt|\lqcd}+e_c \cdot e\right) \bigg] \; ,
\end{multline}
where $T_A$ is the standard nuclear transverse thickness function,
\begin{equation}
T_A(\bt)= \int dz \frac{n}{1+\exp \left[ \frac{\sqrt{\bt^2 + z^2}-R_A}{d} \right]} \; ,
\end{equation}
with $d=0.54\,\mathrm{fm}$ and $R_A=(1.12A^{1/3}-0.86A^{-1/3})\,\mathrm{fm}$. Here $n$ is fixed so that the distribution is normalized to unity. The other parameters in Eq.~(\ref{eq:ica}) take the same values as in the case of a proton target. Because $S$ now depends on $\bt$ we integrate explicitly Eq.~(\ref{eq:dsigma}) over the impact parameter when computing proton-nucleus cross sections. At large impact parameters, the saturation scale of the nucleus falls below the one of the proton~\cite{Lappi:2013zma}. In this region where the nucleus is too dilute for this parametrization to be reliable, we use the proton-proton result scaled such that the nuclear modification factor is unity. We emphasize that in this model, besides the standard Woods-Saxon transverse thickness function $T_A$, no new parameters are introduced when going from proton-proton to proton-nucleus collisions. Using, in contrast to e.g.~\cite{Basso:2016ulb}, these proton and nucleus dipole correlators already used for light hadron~\cite{Lappi:2013zma} and $J/\psi$~\cite{Ducloue:2015gfa,Ducloue:2016pqr} production leads to a rather precise prediction for the nuclear modification factor of forward Drell-Yan production at the LHC as will be shown in the next section.

\section{Results}

\begin{figure}
	\hspace{-1.1cm}
	\includegraphics[scale=1.25]{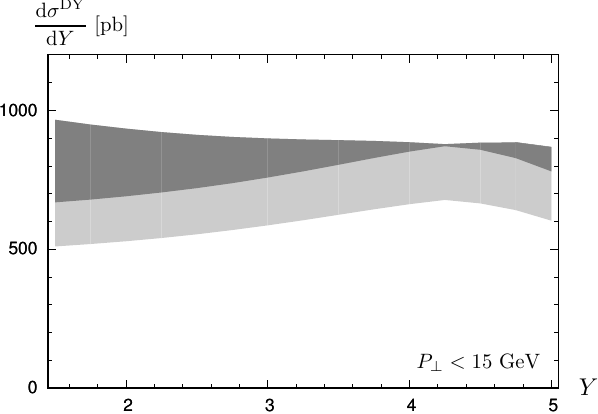}
	\caption{Proton-proton cross section as a function of rapidity at a center of mass energy $\sqrt{s}=8.16$ TeV.}
	\label{fig:sigma_pp_Y}
\end{figure}

In this section we present our results for the cross section and nuclear modification factor of forward Drell-Yan production at the LHC at a center of mass energy $\sqrt{s}=8.16$ TeV. We consider dilepton invariant masses in the range $5 \, \text{GeV} <M<9.25 \, \text{GeV}$. Low invariant masses give access to small $x_2$ values in the target, making saturation effects stronger. This low mass region was shown to be accessible experimentally at LHCb~\cite{LHCb-CONF-2012-013}. The effect of varying the factorization scale $Q$ between $M/2$ and $2M$ when using the definition~(\ref{eq:x2_min}) for $x_2$ is shown as a dark uncertainty band, while a light band shows the same effect when using instead the definition~(\ref{eq:x2_eff}) for $x_2$. We use the leading order MSTW2008 parametrization~\cite{Martin:2009iq} to describe the quark densities in the projectile proton, taking into account the three light flavors.

\begin{figure}
	\includegraphics[scale=1.25]{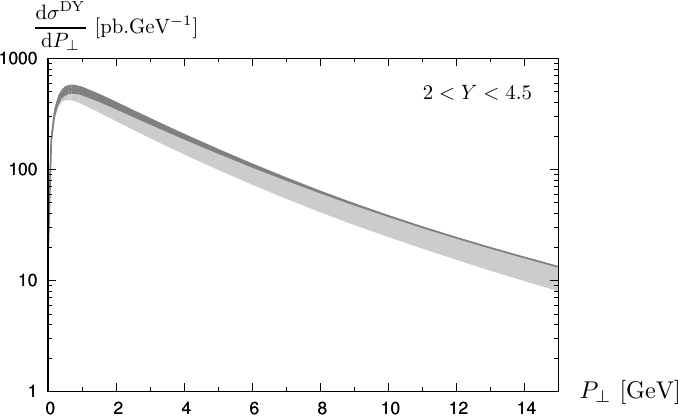}
	\caption{Proton-proton cross section as a function of transverse momentum at a center of mass energy $\sqrt{s}=8.16$ TeV.}
	\label{fig:sigma_pp_pT}
\end{figure}

In Fig.~\ref{fig:sigma_pp_Y} we show the proton-proton cross section as a function of rapidity integrated over $P_\perp$ up to 15 GeV. We note that the formalism used here is not expected to be reliable at high transverse momenta, where a description in collinear factorization would be more suitable. However, since the cross section decreases quickly at large $P_\perp$, this region gives only a small contribution to the total cross section. We observe a rather large uncertainty due both to the choice of $x_2$ and $Q$. In particular, different choices for $Q$ can lead to different trends: while the choice $Q=M/2$ leads to a generally increasing cross section as a function of $Y$, the other extreme choice $Q=2M$ leads to a generally decreasing cross section. This is due to the behavior of quark densities in the projectile. In Fig.~\ref{fig:sigma_pp_pT} we also show the $P_\perp$ spectrum in proton-proton collisions integrated over rapidity in the range $2<Y<4.5$.

\begin{figure}
	\hspace{-1.1cm}
	\includegraphics[scale=1.25]{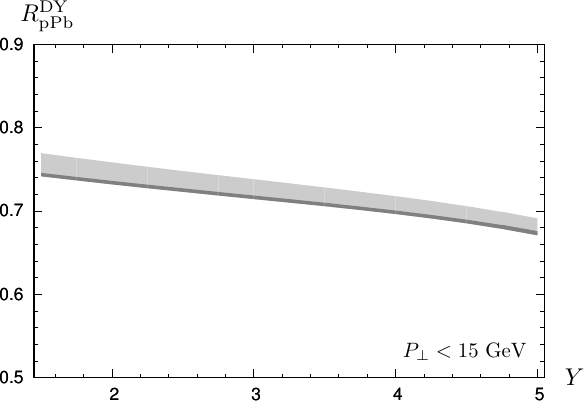}
	\caption{Nuclear modification factor as a function of rapidity at a center of mass energy $\sqrt{s}=8.16$ TeV.}
	\label{fig:RpA_Y}
\end{figure}

\begin{figure}
	\includegraphics[scale=1.25]{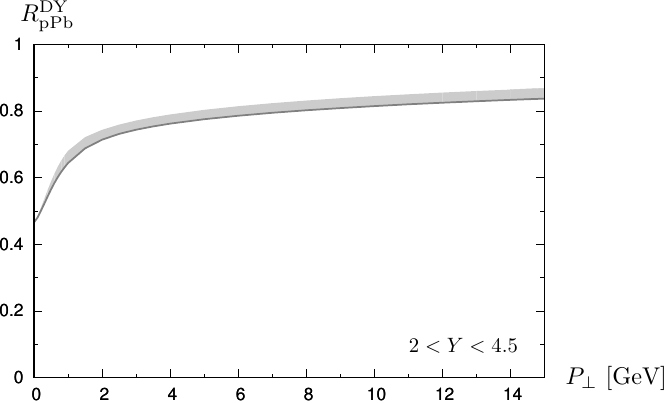}
	\caption{Nuclear modification factor as a function of transverse momentum at a center of mass energy $\sqrt{s}=8.16$ TeV.}
	\label{fig:RpA_pT}
\end{figure}

While the absolute cross section can be quite sensitive to scale variations, the nuclear modification factor is in general a more robust observable. Indeed, normalization uncertainties will cancel to some extent in this ratio defined as
\begin{equation}
R_\text{pA}= \frac{1}{A}\frac{\left . \ud \sigma/ \ud 
	P_\perp \ud Y \right |_\text{pA}}
{\left . \ud\sigma/\ud P_\perp \ud Y \right |_\text{pp}} \; .
\end{equation}
This is indeed the case here, as can be seen from Figs.~\ref{fig:RpA_Y} and~\ref{fig:RpA_pT} where we show the nuclear modification factor for Drell-Yan production as a function of rapidity and transverse momentum respectively. Therefore, this observable could provide an interesting test of the formalism used here and it could for example be measured at LHCb in the near future~\cite{LHCb-PUB-2016-011}.

\begin{figure}
	\includegraphics[scale=1.25]{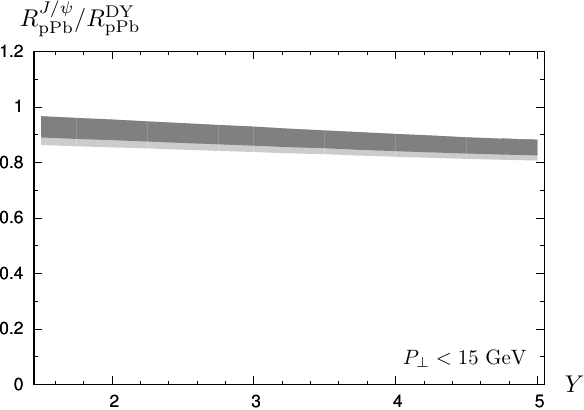}
	\caption{Ratio of the nuclear modification factors of $J/\psi$ and Drell-Yan production as a function of rapidity at a center of mass energy $\sqrt{s}=8.16$ TeV.}
	\label{fig:ratio_Jpsi_DY}
\end{figure}

Beyond the interest for Drell-Yan production itself, the values of the nuclear modification factor presented here can be compared with the results obtained for other processes in the same formalism, such as forward $J/\psi$ production. Indeed, it was recently suggested~\cite{Arleo:2015qiv} that the ratio $R_\text{pA}^{J/\psi}/R_\text{pA}^\text{DY}$ could be used to disentangle between several approaches which are compatible with the rapidity dependence of the nuclear modification of $J/\psi$ production at the LHC. Computing this ratio, using the same dipole correlators as in Refs.~\cite{Ducloue:2015gfa,Ducloue:2016pqr} for consistency, is therefore one of the main objectives of the present work. The variation of this ratio as a function of the rapidity of the lepton pair is shown in Fig.~\ref{fig:ratio_Jpsi_DY}. One can observe that this ratio is rather flat and close to unity, as could be expected from the similar behavior of the nuclear modification factor for Drell-Yan production shown in Fig.~\ref{fig:RpA_Y} and the one obtained in the case of $J/\psi$ production in Refs.~\cite{Ducloue:2015gfa,Ducloue:2016pqr}.

It should be noted that the hadronization of $c\bar{c}$ pairs into $J/\psi$ mesons is not yet fully understood, and that the results for $R_\text{pA}^{J/\psi}$ used here were obtained in Refs.~\cite{Ducloue:2015gfa,Ducloue:2016pqr} by using the color evaporation model. In this model, a fixed fraction of the $c\bar{c}$ pairs produced below the $D$ meson mass threshold hadronize into $J/\psi$ mesons. Therefore, since this fraction is taken to be the same in proton-proton and proton-nucleus collisions, it cancels when calculating the nuclear modification factor and the only uncertainties taken into account when computing $R_\text{pA}^{J/\psi}$ are the variation of the factorization scale and of the charm quark mass (see Refs.~\cite{Ducloue:2015gfa,Ducloue:2016pqr} for more details). Using another mechanism to describe $J/\psi$ hadronization, such as non-relativistic QCD (see Ref.~\cite{Ma:2015sia}), could lead to different results for $R_\text{pA}^{J/\psi}$ and thus also for the ratio $R_\text{pA}^{J/\psi}/R_\text{pA}^\text{DY}$ shown in Fig.~\ref{fig:ratio_Jpsi_DY}.

While the results shown here for $R_\text{pA}^{J/\psi}/R_\text{pA}^\text{DY}$ cannot be directly compared with those shown in Ref.~\cite{Arleo:2015qiv} because of the different center of mass energies and dilepton invariant mass ranges considered, it is interesting to note that the behavior of this ratio as a function of rapidity can be very different depending on the approach followed. In collinear factorization, at leading order $J/\psi$ production probes the gluon density of the target while Drell-Yan production involves the quark distributions. Because the nuclear PDFs are still not yet strongly constrained by data, the predictions for the ratio $R_\text{pA}^{J/\psi}/R_\text{pA}^\text{DY}$ in this approach show a relatively wide spread compatible with values close to unity~\cite{Arleo:2015qiv}, as are the results presented here. The contrast is much more drastic when comparing with the results obtained in the coherent energy loss model~\cite{Arleo:2012hn,Arleo:2012rs}, in which this ratio decreases quickly as the rapidity increases~\cite{Arleo:2015qiv}. Therefore, this observable could help to discriminate between approaches based on the modification of parton densities on one hand and on medium-induced radiation on the other hand. More generally, an accurate measurement of the nuclear modification of forward Drell-Yan production at the LHC would provide valuable information on parton densities at small $x$ in a nucleus which could be used to improve the accuracy of the predictions made either in the color dipole approach or in collinear factorization.

\section{Conclusions}

In this work we have studied forward Drell-Yan production in high energy proton-nucleus collisions at the LHC in the color dipole formalism, using for the description of the dense target the same dipole correlators as in Refs.~\cite{Lappi:2013zma,Ducloue:2015gfa,Ducloue:2016pqr}. In particular, we used the optical Glauber model to obtain the dipole correlator of a nucleus from the one of a proton. This avoids the need to introduce new free parameters to describe a nuclear target. This approach was shown in Refs.~\cite{Lappi:2013zma,Ducloue:2015gfa,Ducloue:2016pqr} to lead to a rather good agreement with experimental measurements of the nuclear modification of forward light hadron and $J/\psi$ production. The comparison of the nuclear modification factors presented here with future measurements would provide an additional test for these correlators, which are assumed to be process-independent. In addition, using the same dipole correlators as in Refs.~\cite{Ducloue:2015gfa,Ducloue:2016pqr} allowed us to compute consistently the ratio $R_\text{pA}^{J/\psi}/R_\text{pA}^\text{DY}$, which was recently proposed as a way to distinguish between several approaches that can describe the nuclear modification of $J/\psi$ production at the LHC~\cite{Arleo:2015qiv}. An experimental determination of this ratio would therefore be extremely valuable to better understand $J/\psi$ suppression in high energy proton-nucleus collisions.

\section*{Acknowledgements}
We thank F.~Arleo for discussions, T.~Lappi for comments on this manuscript and H.~Mäntysaari for providing the dipole cross sections used here.
This work has been supported by the European Research Council, grant ERC-2015-CoG-681707, and by computing resources from 
CSC -- IT Center for Science in Espoo, Finland.

\providecommand{\href}[2]{#2}\begingroup\raggedright\endgroup

\end{document}